# High Quality Single Crystal of Kitaev Spin Liquid Candidate Material RuBr$_3$ Synthesized under High Pressure

Bowen Zhang(张博文)[1], Xiangjun Li(李相君)[1,2], Limin Yan(闫立敏)[1], Wenbo Li(李文博)[1], Nana Li(李娜娜)[1], Jianfa Zhao(赵建发)[3], Xiaobing Liu(刘晓兵)[2], Shun-Li Yu(于顺利)[4], Zhiwei Hu(胡志伟)[5], Wenge Yang(杨文革)[1], Runze Yu(于润泽)[1,*]

[1]Center for High-Pressure Science and Technology Advanced Research, Beijing 100093, China

[2]Laboratory of High-Pressure Physics and Material Science, School of Physics and Physical Engineering, Qufu Normal University, Qufu 273165, China

[3]Institute of Physics, Chinese Academy of Sciences, Beijing 100190, China

[4]National Laboratory of Solid-State Microstructures and Department of Physics, Nanjing University, Nanjing 210093, China

[5]Max-Planck Institute for Chemical Physics of Solids, Dresden 01187, Germany

*Corresponding author: runze.yu@hpstar.ac.cn

**Abstract:**

Kitaev quantum spin liquids have attracted significant attention in condensed matter physics over the past decade. To understand their emergent quantum phenomena, high-quality single crystals of substantial size are essential. Here, we report the synthesis of single crystals of the Kitaev quantum spin liquid candidate RuBr$_3$, achieving millimeter-sized crystals through a self-flux method under high pressure and high temperature conditions. The crystals exhibit well-defined cleavage planes with a lustrous appearance. Transport characterizations exhibit a narrow band-gap semiconducting behavior with 0.13 eV and 0.11 eV band-gap in *ab* plane and along *c* axis, respectively. Magnetic measurement shows a transition to antiferromagnetic (AFM) state at approximately 29 K both in *ab* plane and along *c* axis. Notably, the Néel temperature increases to 34 K with an applied magnetic field of up to 7 T in the *ab* plane, but without any change along *c* axis. The large size and high quality of RuBr$_3$ single crystals provide a valuable platform for investigating various interactions, particularly the Kitaev interaction, and for elucidating the intrinsic physical properties of Kitaev quantum spin liquids.

**Introduction:**

Owing to the exotic emergent state of matter exhibiting Majorana fermion and gauge flux excitations, the Kitaev quantum spin liquids have attracted considerable interest in last decade[1-10]. Unlike quantum spin liquids arising from geometrical frustrated spin arrangements [11-18], the Kitaev quantum spin liquid state can be realized in honeycomb structure with strong spin-orbital coupling (SOC). In such structures, the coupling between neighboring spins is highly anisotropic with bond-dependent spin interactions and induces frustration of spin configuration on a single site[2-7, 19]. Recently several Kitaev-type quantum spin liquid candidates have been synthesized in 4d and 5d compounds with strong SOC, such as $A_2IrO_3$ (A = Na or Li) and α-$RuCl_3$[2-4, 18-27], in which the 4d ($Ru^{3+}$) or 5d ($Ir^{4+}$) ions form a Mott insulating state on a honeycomb lattice, and the localized electrons exhibit an effective spin $J_{eff}$=1/2 due to the strong SOC effect.

Among the Kitaev-type quantum spin liquid candidates, the most promising one is α-$RuCl_3$. This compound possesses a layered structure with edge-sharing $RuCl_6$ octahedra arranged in a honeycomb lattice plane. The $Ru^{3+}$ ions in α-$RuCl_3$ have the same $t^5_{2g}$ configuration as $Ir^{4+}$ ions in the iridates, and $J_{eff}$=1/2 Mott insulating state was realized with the assistance of weak SOC effect [28]. Kitaev interaction has been identified in this compound and spin-liquid physics signatures, like broad continuum of magnetic excitations were identified both in Raman scattering[4] and inelastic neutron scattering measurements[3, 6, 8], despite the presence of a zigzag antiferromagnetic order around 7-14 K. Moreover, this zigzag antiferromagnetic order can be suppressed under magnetic field or pressure as it is close to the Kitaev quantum spin liquid phase in the phase diagram[7, 29-32]. The interactions of both Kitaev and no-Kitaev in α-$RuCl_3$ can be described by an extended model named J-K-Γ model, in which K, J and Γ represent the Kitaev interactions, Heisenberg interaction and off-diagonal interaction respectively [33-35]. However, the absence of analogous materials for α-$RuCl_3$ makes it challenging to tune the parameters J, K, and Γ to push the ground state closer to the Kitaev limit. Additionally, α-$RuCl_3$ single crystals often exhibit a large number of stacking faults in the crystal structure, leading to multiple magnetic transitions and complicating the investigation of intrinsic magnetic interactions. Thus, finding new quantum spin liquid analog with α-$RuCl_3$ has become a crucial issue.

Recently a Kitaev spin-liquid candidate material $RuBr_3$ with a honeycomb lattice has been synthesized under high pressure and high temperature[36]. Like its sister compound α-$RuCl_3$, $RuBr_3$ is an insulator with a Zigzag antiferromagnetic order at around 34 K, but with dominant antiferromagnetic interactions and a direction of the zigzag-ordered moment different from α-RuCl₃, indicating closer proximity to the pure Kitaev state[36]. Moreover, the replacement from Cl to Br allows Kitaev and non-Kitaev interactions can be tuned in these ruthenium trihalides by varying the ligand sites, which provides a new platform for exploring Kitaev spin liquids property. However, the ratio of different interactions, the magnetic behaviors under high pressure or magnetic field, even spin liquid physics signatures, like a broad continuum of magnetic excitations are still elusive. According to the previous works on α-RuCl₃, most intrinsic physical property characterizations are highly depending on large size and high quality single crystals for its strong anisotropic property[2-10]. Therefore, synthesizing high-quality single crystals is crucial. However, this honeycomb structure can only be stabilized under high pressure and high temperature, and it's extremely challenge to grow large size single crystal under high pressure, especially the Br element is very reactive at high temperatures. So far there is still no report of large size (up to millimeter) single crystal growth on $RuBr_3$.

In this paper, we report the design of special assemble for high pressure synthesis and successfully growth of $RuBr_3$ single crystal up to millimeter under high pressure and high temperature. Single crystal diffraction indicated it crystallizes to *R*-3 space group (No.148), which is consistent with the result from powder diffraction. Transport and magnetic properties along *ab* plane and *c* axis were carefully investigated.

**Experimental:**

Singel crystals of $RuBr_3$ were grown using high pressure cubic anvil apparatus. The precursor of $RuBr_3$ was purchased from Alfa-Aesar Chemical Company with high purity (Ru 29% min. in weight). The precursor was sealed into our designed assemble, then placed into the center of the high-pressure cells and compressed up to 4 GPa. The temperature was first quickly increased to 800°C with rate 200 °C /min and maintained at that temperature for 2h to form the polycrystalline sample with honeycomb structure. Then the temperature was increased to 1200°C and held for 2h, followed by a decrease to 800°C in 6 hours. Finally, the sample was quenched to room temperature before releasing pressure. X-ray diffraction was conducted at room temperature, using an X-ray diffractometer (D/max 2500V) with Cu *K*α radiation at operating voltage and current of about 40 kV and 150 mA, respectively. The magnetic properties were measured using the MPMS3 of

Quantum Design under a DC mode. Scanning Electron Microscope (SEM) was carried out by JEOL, JSM-7000F. Chemical compositions of single crystals were determined using a scanning electron microscope with an energy dispersive X-ray analyzer. Transport measurement was performed using four-probe method. The heat capacity measurements were carried out on a Physical Property Measurement System.

**Results and Discussion:**

High pressure single crystal growth is a very challenge task. The most difficult step is sealing the precursor materials with full density to ensure they remain clean and intact in the high-temperature melting state. Typically, to avoid the contamination from the sample holder (such as hBN and carbon heater) or capsule, and maintain a suitable atmosphere, precious metals like Pt and Au are used to seal the sample during high-pressure experiments, like the experiment in cuprate superconductor of $Ca_{2-x}Na_xCuO_2Cl_2$ [37-38]. The reason for choosing precious metals is their chemical stability under extreme conditions (such as high pressure and high temperature) and their lack of reaction with precursors. In this project, we initially chose a Pt foil with a thickness of 0.03 mm as the sample capsule to seal the sample. The sample capsule was loaded into an hBN tube, assembled into a carbon tube (See left part of Fig. 1), and placed into the center of pressure media (usually pyrophyllite) with high pressure and temperature treatment at 4 GPa and 1200°C for 30 minutes followed by annealing in 800°C in 3 hours. Bright spots in the sample and nearly layered grains were observed after crushing the high-pressure-treated sample and removing the Pt capsule, indicating that some single crystals had been obtained. However, the crystal quality was poor and can't get larger size sample. Moreover, we also noticed that the bottom of the Pt capsule covered some of materials with yellow color and no bright spots can be seen at that part of sample, indicating that the Pt capsule alone could not completely seal the sample during the experiment. The same phenomenon can also be observed even when using a double-layer Pt capsule. We believe there have been reaction between Pt capsule and $RuBr_3$. Therefore, the environment for single crystal growth was not stable and was detrimental to the precipitation and growth of single crystals. To overcome this deficiency, we designed a new assemble to seal the sample. First, we sealed sample in Pt capsule and insert it into hBN capsule and then we sealed the hBN capsule with a new Pt capsule (See right part of Fig.1). Using this assemble and the same process of heating mentioned in the experimental part, we finally got high quality single crystals of $RuBr_3$ up to millimeter size.

The high-quality single crystals, with typical sizes up to 0.8 mm×0.5 mm×0.05 mm, were obtained after breaking the reaction samples. They exhibit a dark color with a shiny surface (Inset a of Fig. 2), leaving a flat cleavage *ab* plane surface. These properties are consistent with previous result showing semiconducting behavior with energy 0.21 eV and two-dimensional structure. The SEM image shows obvious two-dimensional properties and well-defined easy cleavage plane (Inset b of Fig 2). The EDX results show the composition of Ru and Br are 74.6% and 25.4% in mole ratio, which is highly consistent with that nominal composition. X-ray diffraction measurement with the incident X-ray along the *c* axis indicated that only sharp peaks along (0 0 3*l*) could be observed, demonstrated high quality crystallization and *c* axis orientation in $RuBr_3$ single crystals.

Figure 3 shows the temperature dependence of the resistivities for $RuBr_3$ along *ab* plane (Fig. 3a) and *c* axis (Fig. 3b) respectively. Both resistivities demonstrate semiconducting behaviors over the entire temperature range. Moreover, both data show thermally activated temperature dependence, which is consistent with a strongly spin-orbital-coupled Mott insulator as reported in previous work [36]. The activation energy in the *ab* plane is 0.13 eV and along *c* axis is about 0.11 eV (See inset of Fig. 3). Both values are smaller than those from polycrystalline samples. Typically, in two-dimensional materials, the resistivity shows strong anisotropy[39-40]. For example, the resistivity along *ab* plane is almost three order magnitude larger than that along the *c* axis for the counterpart $α-RuCl_3$. As to $RuBr_3$, the resistivities along *ab* plane and *c* axis is $10^2$ Ω.cm and $10^{-2}$ Ω.cm at room temperature. That means $RuBr_3$ also exhibits very strong anisotropy in the resistivity, even stronger than that in $α-RuCl_3$. This result also confirmed the deduction that the electronegativity of Br is smaller than that of Cl, making the electronic structure of $RuBr_3$ much more anisotropic than that of $α-RuCl_3$, which might be the origin of the higher $T_N$ in $RuBr_3$[36].

Next, we focus on the magnetic property of $RuBr_3$ single crystal. Fig. 4a and 4b show the temperature dependence of the magnetic susceptibilities under different magnetic fields applied along *ab* plane and *c* axis. Here we also use the notation $χ_c$ to denote susceptibility measured with field applied perpendicular to the honeycomb plane, and $χ_{ab}$ for susceptibility measured with in-plane field. Both the magnetic susceptibility data $χ_{ab}$ and $χ_c$ show broad peak around 60 K which is similar with that observed in the polycrystalline sample, and can be ascribed as the development of antiferromagnetic correlations as reported in previous work[36]. A kink was observed around 29 K for $χ_{ab}$ under 1 T, indicating the formation of AFM order. The $dχ_{ab}/dT$ indicated a sharp peak at about

29 K (See inset of Fig. 4a), which was attributed to the formation of long-range antiferromagnetic order. This value is slightly smaller than that observed in polycrystalline sample. Since magnetic field can completely suppress magnetic order when applied along *ab* plane in α-RuCl$_3$[7], we also focus on the field dependent magnetism transition Neel temperature (the peak position of d$\chi_{ab}$/dT, see inset of Fig. 4a). It can be seen clearly that the Neel temperature increases with the increasing of external magnetic field, reaching a transition temperature of 34 K at 7 T. As the tilt angle of the magnetic moment from the honeycomb plane is α = 64° in RuBr$_3$, which is larger than that in α-RuCl$_3$(α = 32–35°) and makes the magnetic moment closer to that *c* axis, we speculate *c* axis may be the easy suppressed axis in RuBr$_3$. Surprisingly, the AFM transition temperature has nearly no change under high magnetic field along *c* axis (see inset of Fig. 4b). Meanwhile, we notice that the $\chi_{ab}$ is nearly comparable with that $\chi_c$, which is also different from that α-RuCl$_3$, in which $\chi_{ab}/\chi_c$ ~ 8 [41]. Kim *et al.* found the great magnetic anisotropy of α-RuCl$_3$ originated from the large symmetric off-diagonal exchange interaction (referred to Γ term) [38]. Based on this result, we can see the ratios of K, J, Γ might be largely different in RuBr$_3$ compared to α-RuCl$_3$. Field-dependent magnetizations (M(H)) under magnetic fields applied along *ab* plane and *c* axis at 5 K all show liner behavior, indicating no magnetic order at low temperature, consistent with antiferromagnetic behavior.

Fig. 4c and 4d show the temperature dependence of inverse magnetic susceptibilities of 1/$\chi_{ab}$ and 1/$\chi_c$. All the data show linear behaviors between 200-300 K, indicating they follow the Curie-Weiss behavior $\chi=C_{CW}/(T-\theta_{CW})$, in which $C_{CW}$ and $\theta_{CW}$ are Curie constant and Curie-Weiss temperature. The Curie-Weiss temperatures are around $\theta_{ab} \approx$ -42 K and $\theta_c \approx$ -290 K under different magnetic fields, demonstrated the magnetic interactions of RuBr$_3$ are all AFM both for the field applied perpendicular and parallel to the honeycomb plane. This phenomenon is different from that α-RuCl$_3$, which shows ferromagnetic and antiferromagnetic interactions for the susceptibility measured with field applied in-plane and perpendicular to the honeycomb plane, respectively[42, 43]. The effective magnetic moments are about 2.3$\mu_B$ and 3.4$\mu_B$ for $\chi_{ab}$ and $\chi_c$. They are all larger than the spin-only value of 1.73$\mu_B$ for the low-spin state (S = 1/2) of Ru$^{3+}$, which was ascribed from the contribution of the orbital moment, similar to α-RuCl$_3$[43]. According to previous report[33], in the J-K-Γ model, the Curie-Weiss temperature anisotropy satisfies ($\theta_c-\theta_{ab}$)/($\theta_C+2\theta_{ab}$)=Γ/(3J+K) and there are relationships Γ/(3J + K) ~ ∞ for α-RuCl$_3$ [10] and Γ/(3J + K) ~ −0.3 for Na$_2$IrO$_3$[18]. However, the Γ/(3J + K) is about 0.7 in RuBr$_3$, indicating that the interactions in RuBr$_3$ may different from that $\alpha$-RuCl$_3$ and Na$_2$IrO$_3$. This may suggest the presence of more complex interactions in RuBr$_3$, such as ring exchange interactions [44], due to its smaller charge gap. This result also coincides with the magnetic anisotropy mentioned above. We also measured the specific heat for single crystal (See Figure 5). Unfortunately, no obvious jump was observed around 30 K for single crystals (See inset of Figure 5). In fact, the jump of specific heat reported in the previous work was also very weak[36]. According to the previous work, the temperature dependence of the magnetic susceptibility showed a broad peak around 60 K, which is also observed in our work based on single crystal, indicating the formation of antiferromagnetic correlation[36]. Thus, the magnetic entropy was released in a broad temperature range instead of one point, so there was no jump in the specific heat data.

Kitaev spin liquid state for 4d$^5$ configuration of Ru$^{3+}$ has $J_{eff}$ =1/2 ground state considering both a relatively large SOC and pure $t_{2g}^5$. The mixing of some $e_g$ character into the $t_{2g}$ manifold will break down $J_{eff}$ =1/2. It was found that the large SOC of 5d state can mix $e_g$ character into the $t_{2g}$ manifold leading to a deviation from $J_{eff}$ =1/2 for Ir$^{4+}$ oxide[45]. For 4d oxide $e_g$-$t_{2g}$ mixing induced by SOC is small. The $e_g$-$t_{2g}$ mixing can occur by the presence of a trigonal crystal field ($D_{trig}$), which is large or comparable to SOC splitting. However, the actual $e_g$-$t_{2g}$ mixing induced by trigonal crystal field is small for $\alpha$-RuCl$_3$ and RuBr$_3$. For example, the Ru $L_3$-edge resonant inelastic x-ray scattering experiments indicated that the $t_{2g}$ splitting by spin-orbit coupling is about 225 meV[46], while $D_{trig}$ for $\alpha$-RuCl$_3$ and RuBr$_3$ could not be observed within experimental resolution of ~80 meV. In fact $D_{trig}$ is only about 12 meV for $\alpha$-RuCl$_3$[47]. Therefore, the trigonal crystal field distortion is expected to be negligible for RuBr$_3$. The different physical properties between RuBr$_3$ and $\alpha$-RuCl$_3$ can be attributed to a relatively large Ru4d-Br4p hybridization in RuBr$_3$ as compared with Ru4d-Cl3p hybridization.[45].

Theorist proposed that the "pseudomoment" direction, related to the Kitaev interaction in Ref.[48], is not exactly the same as the moment angle ($\alpha$) experimentally obtained by neutron scattering. The angle α is influenced by the anisotropy of the g-factors. Thus, it is very desirable to carefully examine the magnetic properties along different directions in the future work to explore the possibility of observing a pure Kitaev interaction. Moreover, since the maximum magnetic field applied to the sample is currently limited to 7 T, the effects of higher magnetic fields need further investigation to determine the potential for suppressing magnetic order.

**Conclusion:**

We synthesized single crystals of Kitaev-magnetism compound RuBr$_3$ under high pressure and high temperature. Transport measurements reveal highly anisotropic properties. Magnetic characterization shows AFM around 29 K both along *ab* plane and *c* axis. Furthermore, the AFM transition temperature along *ab* plane from approximately 29 K to 34 K as the magnetic field increases from 1 T to 7 T, while it remains robust along the c-axis. The high-quality and large size of RuBr$_3$ single crystal provides an ideal platform to further investigate the different magnetic-interaction and electronic structure using spectroscopic techniques, such as inelastic neutron diffraction, NMR and some transport measurement like heat transport. Additionally, the physical properties of RuBr$_3$ based on the single crystal may offer insights into understanding the differences between its sister compound α-RuCl$_3$ and provide a pathway to establish the structure and different magnetic interactions, potentially guiding the search for a pure Kitaev quantum spin liquid state.

**Acknowledgments:**

This work was supported by National Key Research and Development Program of China (Grant No. 2023YFA1406000, 2022YFA1403800), the National Natural Science Foundation of China (Grant Nos. 12474002, 22171283, 22203031, 12434005, 12204515 and 12074175). J.F.Z. acknowledges support of the Young Elite Scientists Sponsorship Program by CAST (Grant No. 2022QNRC001). We thank Prof. Jinsheng Wen for the helpful discussion.

**Figure captions:**

**Figure 1**: Cell assembles for $RuBr_3$ single crystals growth under high pressure.

**Figure 2**: X-ray diffraction pattern of $RuBr_3$ single crystal along the *c* axis. Inset (a) Photograph of $RuBr_3$ single crystal. Inset (b) SEM image of $RuBr_3$ single crystal.

**Figure 3**: Temperature dependent resistivities of $RuBr_3$ single crystals along (a) *ab* plane and (b) *c* axis. The insets show the Arrhenius plot, with red lines indicating the fitting results.

**Figure 4**: (a, b) Temperature dependence of magnetic susceptibilities for $RuBr_3$ single crystal under magnetic field of $\mu_0H$ = 1, 3, 5, 7 T applied in and perpendicular to the honeycomb plane. The insets show the temperature differential of $d\chi_{ab}/dT$ and $d\chi_c/dT$ under magnetic field of $\mu_0H$ = 1, 3, 5, 7 T. The peak positions of $d\chi_{ab}/dT$ (indicating the Neel temperature) increase with increasing external magnetic field, while there is no change for $d\chi_c/dT$. (c, d) Field-dependent magnetization of $RuBr_3$ single crystals under magnetic field applied in and perpendicular to the honeycomb plane at 5K. (e, f) Inverse of magnetic susceptibility under a magnetic field of $\mu_0H$ = 1, 3, 5, 7 T for $\chi_{ab}$ and $\chi_c$ respectively. The bule dash lines show the Curie-Weiss law fitting results.

**Figure 5**: Temperature dependence of specific heat for $RuBr_3$ single crystals. The inset shows the enlarged part around 20-40 K.

**Figure 1**

Sample
hBN
Pt capsule
Graphite heater

**Figure 2**

(003)
(006)
(009)
(0012)

Intensity (a.b.)
2 theta (deg.)

(a) 1 mm
(b)

**Figure 3**

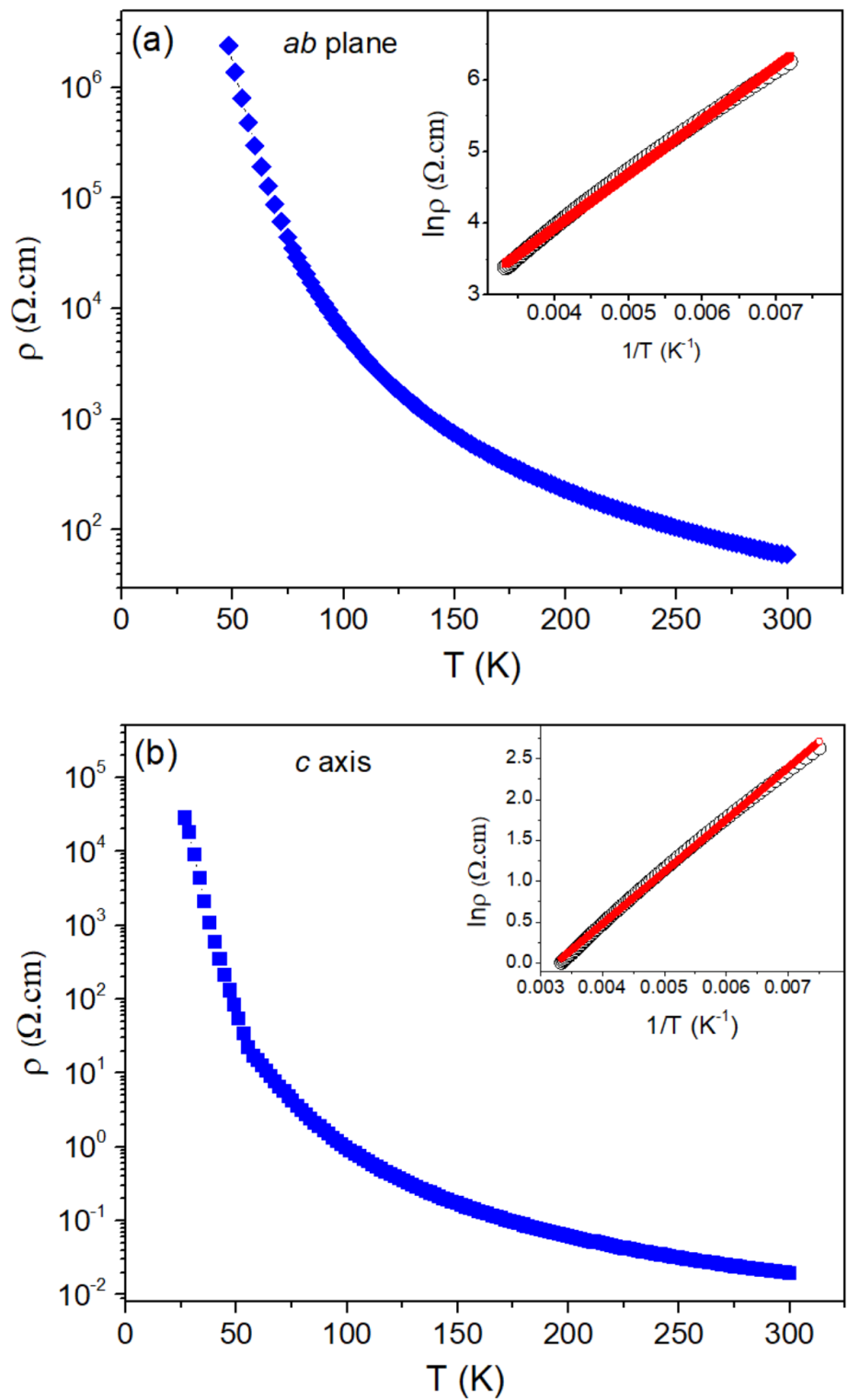

**Figure 4**

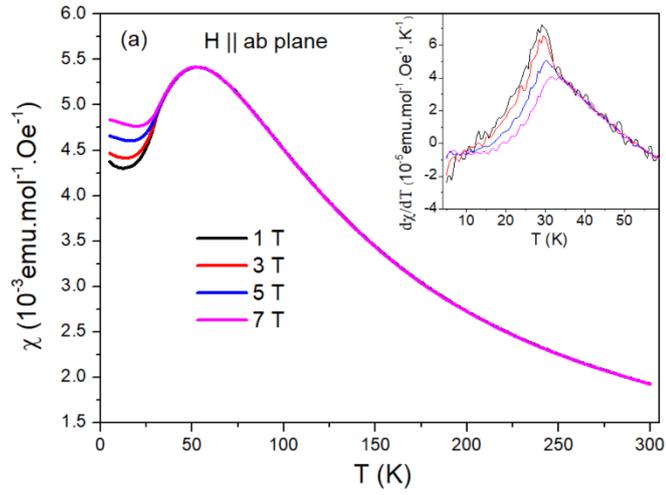
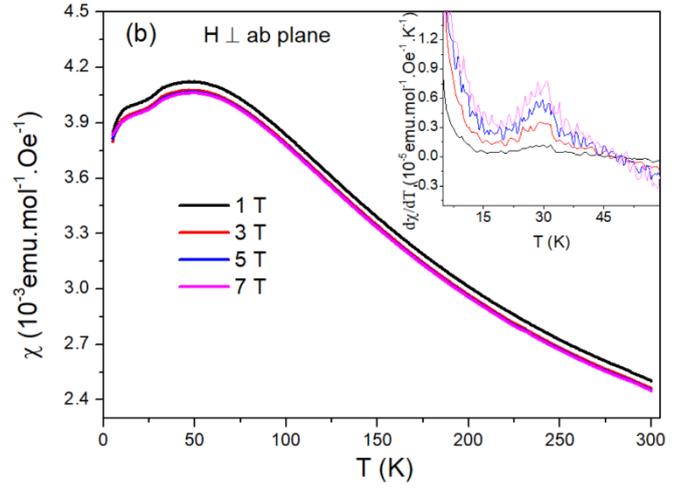
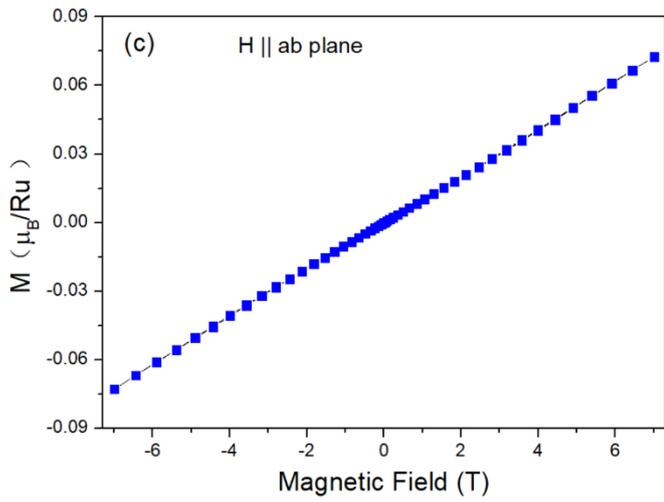
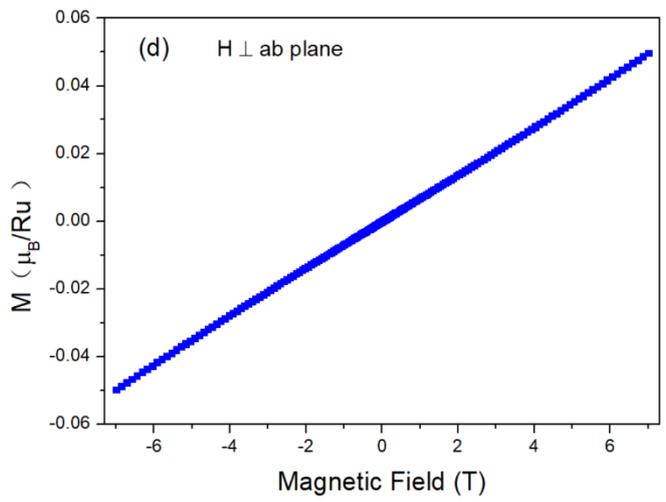
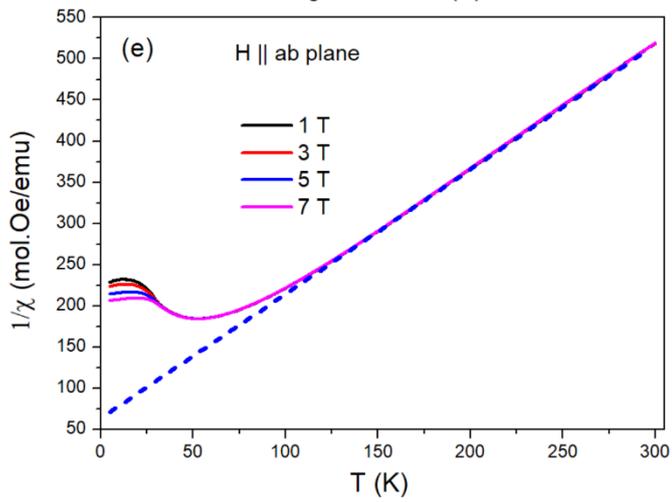
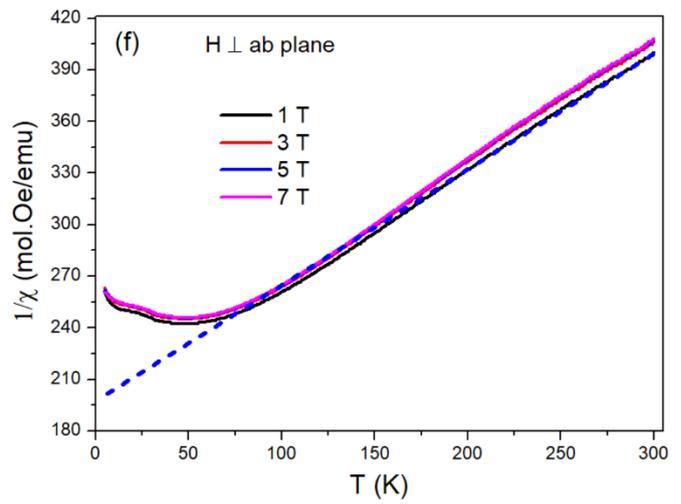

**Figure 5**

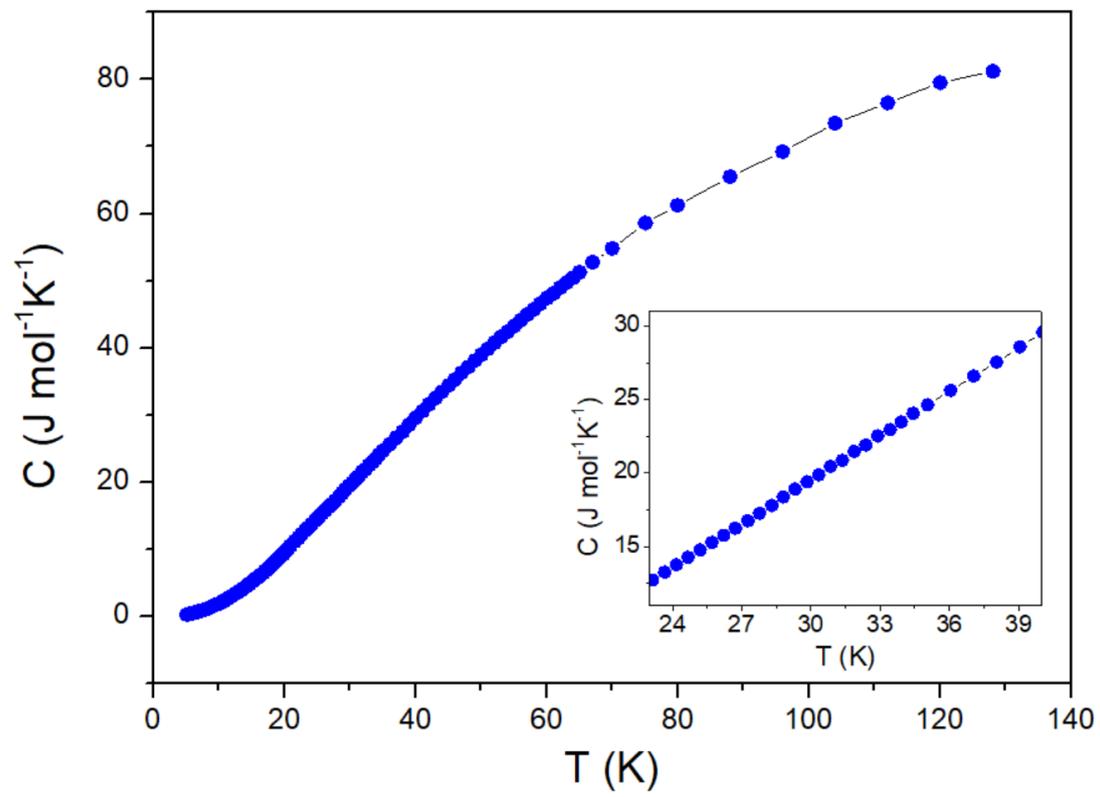